\documentclass[aps,prl,twocolumn]{revtex4-2}
\usepackage{amssymb}
\usepackage{amsmath,bm}
\usepackage{graphicx,color}
\usepackage{bbm}
\usepackage{multirow}
\newcommand{\B}[1]{{\bm{#1}}}%% Bold Roman & Greek Lower & Upper Case

\usepackage{hyperref}

\begin{document}

\title{Density of Quasi-localized Modes in Glasses: where are the Two-Level Systems?}
\author{Avanish Kumar$^1$, Itamar Procaccia$^{1,2}$ and Murari Singh$^3$}
\affiliation{$^1$Department of Chemical Physics, The Weizmann Institute of Science, Rehovot 76100, Israel. \\$^2$  Center for OPTical IMagery Analysis and Learning, Northwestern Polytechnical University, Xi'an, 710072 China. \\$^3$ Dept. of Physics, University of Texas, Austin Tx. 78712 }

\begin{abstract}
The existence of a constant density of two-level systems (TLS) was proposed as the basis of some intriguing universal aspects of glasses at ultra-low temperatures. Here we ask whether their existence is necessary
for explaining the universal density of states quasi-localized modes (QLM) in glasses at ultra-low temperatures. A careful examination of the QLM that exist in a generic atomistic model of a glass former reveals
at least two types of them, each exhibiting a different density of states, one depending on the frequency
as $\omega^3$ and the other as $\omega^4$. The properties of the glassy energy landscape that is
responsible for the two types of modes is examined here, explaining the analytic feature responsible for the creations of (at least) two families of QLM's. Although adjacent wells certainly exist in the complex energy landscape of glasses, doubt is cast on the relevance of TLS for the universal density of QLM's. 
\end{abstract}

\maketitle

{\bf Introduction}: Low frequency vibrations in solids are delocalized Debye modes. In contradistinction, athermal glasses exhibit an excess of low-frequency modes which have attracted considerable attention over the  years due to their important contributions to
the Boson peak, to enhanced heat capacity and to plastic responses \cite{81PA,88Kli}. For many years the unusual universal features of glasses at ultra-low temperatures,
were explained by a ``tunneling model" assuming the existence of a constant distribution of two-level tunneling states (TLS) \cite{72AHV,72Phi}. In classical mechanics such two-level systems are envisaged as vibrations over a `double well' neighborhood on the energy landscape, with small barriers in between the two wells. While there
are no doubts about the existence 
of such close-by wells on the very complex energy landscape of generic glasses \cite{93HS,15GL}, there are questions about
their relevance. Indeed, 
doubts about the importance of TLS were repeatedly voiced \cite{88YL,14PRRR,17ALMCR,20KSBRZ}.  In this Letter we ask a very specific question, i.e. whether the existence of double-well structures is important for explaining the universal density of states of QLM's in generic glasses. To this aim we employ a generic atomistic model of a glass former for which the low-frequency vibrations and their analytic structure can be computed explicitly. The model allows a direct demonstration of the universality of the density of states of quasi-localized modes, but the analysis shows explicitly that TLS play no role in establishing this universal behavior.

On purely theoretical grounds it was predicted for more than thirty years now \cite{83KKI,87IKP,91BGGS,03GC,03GPS,07PSG,14SBG} that in athermal amorphous solids the access modes should exhibit a density of states $D(\omega)$ with a universal dependence on the frequency $\omega$, i.e.
\begin{equation}
	D(\omega)\sim \omega^4 \quad \text{in all dimensions} \  .
	\label{dof}
\end{equation}
A successful model for the justification of this result is the so-called ``soft potential" model \cite{91BGGS} which was presented as an extension of the TLS model. To motivate the analysis presented below it is worthwhile to present the thinking behind this model that leads to Eq.~(\ref{dof}).

To start, we recall that ``modes" in amorphous solids emerge in the harmonic approximation. Consider a glassy system of $N$ particles in a volume $V$ at a temperature $T=0$;  The Hamiltonian of the system is denoted $U(\B r_1,\B r_2,\cdots \B r_N)$ and the Hessian matrix is defined as
\begin{equation}
\B H_{ij} \equiv \frac{\partial^2 U(\B r_1,\B r_2,\cdots \B r_N)}{\partial \B r_i \partial \B r_j} \ .
\label{defhes}
\end{equation}
The ``vibrational modes" at $T=0$ are nothing but the eigenfunctions of the Hessian,
denoted below as $\B \Psi^{(k)}$, each associated with an eigenvalue $\lambda_k$. If the system is perturbed in direction of one pure eigenfunction it oscillates indefinitely
with frequency $\omega_k$, $\omega_k^2 = \lambda_k$. Since the Hessian is real and symmetric the eigenvalues and the associated frequencies are all real. When the system is stable the eigenvalues are  all positive with the exception of Goldstone modes associated with symmetries for which $\lambda_k=0$. 

The derivation of Eq.~(\ref{dof}) goes beyond the harmonic approximation. Probably the clearest derivation was presented by Gurarie and Chalker \cite{03GC}. It starts by denoting the position of
the particles as $\B q\equiv \{\B r_i\}_{i=1}^N$. Next assume that the energy landscape has minimum at a phase point $\B q_0$. Consider then a close-by point $\B q_A\ne \B q_0 $ which for convenience is chosen
to be $\B q_A=0$; the potential energy is measured from this point, i.e. $U(\B q_A)=0$. Finally, assume that there exists a QLM whose vibrations are taking place along a `reaction coordinate' $s$ with the energy expanded around our reference point to fourth order:
\begin{equation}
	U(s)=\sum_{n=1}^4 \frac{a_n}{n!} s^n \ .
	\label{ass1}
\end{equation}
We recall that the configuration space is multi-dimensional, and it is therefore crucial to specify the direction of this `reaction coordinate". Different expansions are obtained in different directions. 

Since such QLM's are distributed in phase-space we can have many such local expansions, 
and it is {\em assumed} that due to the glassy disorder the coefficients $a_n$ are quasi-random, with a probability distribution function (PDF) $P(a_1,a_2,a_3,a_4)$ that is finite and smooth. To proceed, consider next a similar
expansion but now around the minimum in the potential $\B q_0$. Since this is a minimum the expansion to the same order reads
\begin{equation}
	U(s)=\sum_{n=2}^4 {b_n} \Big[\frac {(s-s_0)^n}{n!} - \frac{(-s_0)^n}{n!}\Big] \ ,
	\label{ass2}
\end{equation}
where $s_0$ is the value of the {\em assumed} ``reaction coordinate" at $\B q_0$. Finally
with the same logic one assume that there exists a pdf $P(q_0,b_2,b_3,b_4)$. Recall that by definition $b_2=\lambda=\omega^2$, the eigenvalue of the eigenfunction of the Hessian associated with this ``reaction coordinate". 

The next step in the argument is exact, being simply a consequence of change of variables. Demanding that
\begin{eqnarray} 
P(q_0,b_2,b_3,b_4) &=& |J| P(a_1,a_2,a_3,a_4)\ , \nonumber \\
J &\equiv& \frac{\partial (a_1,a_2,a_3,a_4)}{\partial(q_0,b_2,b_3,b_4)} \ ,
\label{demand}
\end{eqnarray}
After little algebra \cite{03GC} one obtains the result
\begin{equation}
P(q_0,b_2,b_3,b_4) = |b_2| P(a_1,a_2,a_3,a_4) \ .
\label{prob}
\end{equation}
If one can now integrate out $q_0, b_3$ and $b_4$ without restrictions (assuming smoothness
and continuity of our PDF), one ends up with 
\begin{equation}
P(b_2)\propto b_2 \ ,  \quad \text{or equivalently}, ~D(\omega)\propto \omega^3 \ .
\end{equation}
To obtain Eq.~(\ref{dof}) one needs to add another crucial assumption, i.e that 
the minimum in question is {\em locally lowest}, or, in other words, if there is another
minimum in the quartic expansion (\ref{ass2}), it is a higher minimum than the one
around which we expand. This is where the picture of double minimum sneaks in, and
we will argue below that it is neither required nor supported by simulation data.
The condition for being at a lower minimum is 
\begin{equation}
|b_3|\le \sqrt{3b_2b_4} \ .	
\label{ineq}
\end{equation}
Integrating out $b_3$ from Eq.(\ref{prob}) with this restriction in mind leads to
\begin{equation}
	P(b_2)\propto b_2^{3/2} \ ,  \quad \text{or equivalently}, ~D(\omega)\propto \omega^4 \ .
\end{equation}
In the rest of this letter we will make these considerations concrete in the context
of a typical model of a glass former, and present a critical assessment of the
assumptions made.

{\bf Model of glass former}: We use here a standard poly-dispersed model of $N=4000$ particles in two dimensions, with density $\rho=1$ in a square box of area $A=4000$.
The units of length is $\sigma_{\rm min}$ as defined below. The binary interactions are
\begin{eqnarray}
	&&\phi(r_{ij}) = \epsilon\left(\frac{\sigma_{ij}}{r_{ij}}\right)^{12} +C_0 +C_2\left(\frac{r_{ij}}{\sigma_{ij}}\right)^2+C_4\left(\frac{r_{ij}}{\sigma_{ij}}\right)^4\nonumber \\&& \epsilon\!=1, C_0\!=\!-1.92415, C_2\!=\!2.11106, C_4\!=\!-0.591097 \ . \nonumber
\end{eqnarray}
The unit of energy will be $\epsilon$ and Boltzmann's constant will be unity.
The interaction length was drawn from a probability distribution $P(\sigma)
\sim 1/\sigma^3$ in a range between $\sigma_{\rm min}$ and $\sigma_{\rm max}$:
\begin{eqnarray}
	&&\sigma_{ij} =\frac{\sigma_i+\sigma_j}{2}\Big[1-0.2\Big|\sigma_i-\sigma_j\Big|\Big],\nonumber\\ 
&&\sigma_{\rm max} =1.45/0.9\ , \sigma_{\rm min}=\sigma_{\rm max}/2.219 \ .
\end{eqnarray}
The parameters are chosen to avoid crystallization and to allow enough smooth derivatives of the Hamiltonian. The system is thermalized at $T=0.1$ using Swap Monte Carlo and then 
cooled down to $T=0$ using conjugate gradient methods.
%%%%%%%%%%%%%%%%%%%%%%%%%%%%%%%%%%%%%%%%%%%%%%%%%%%%%%%%%%%
\begin{figure}
	\hskip -0.5 cm
	\includegraphics[width=0.50\textwidth]{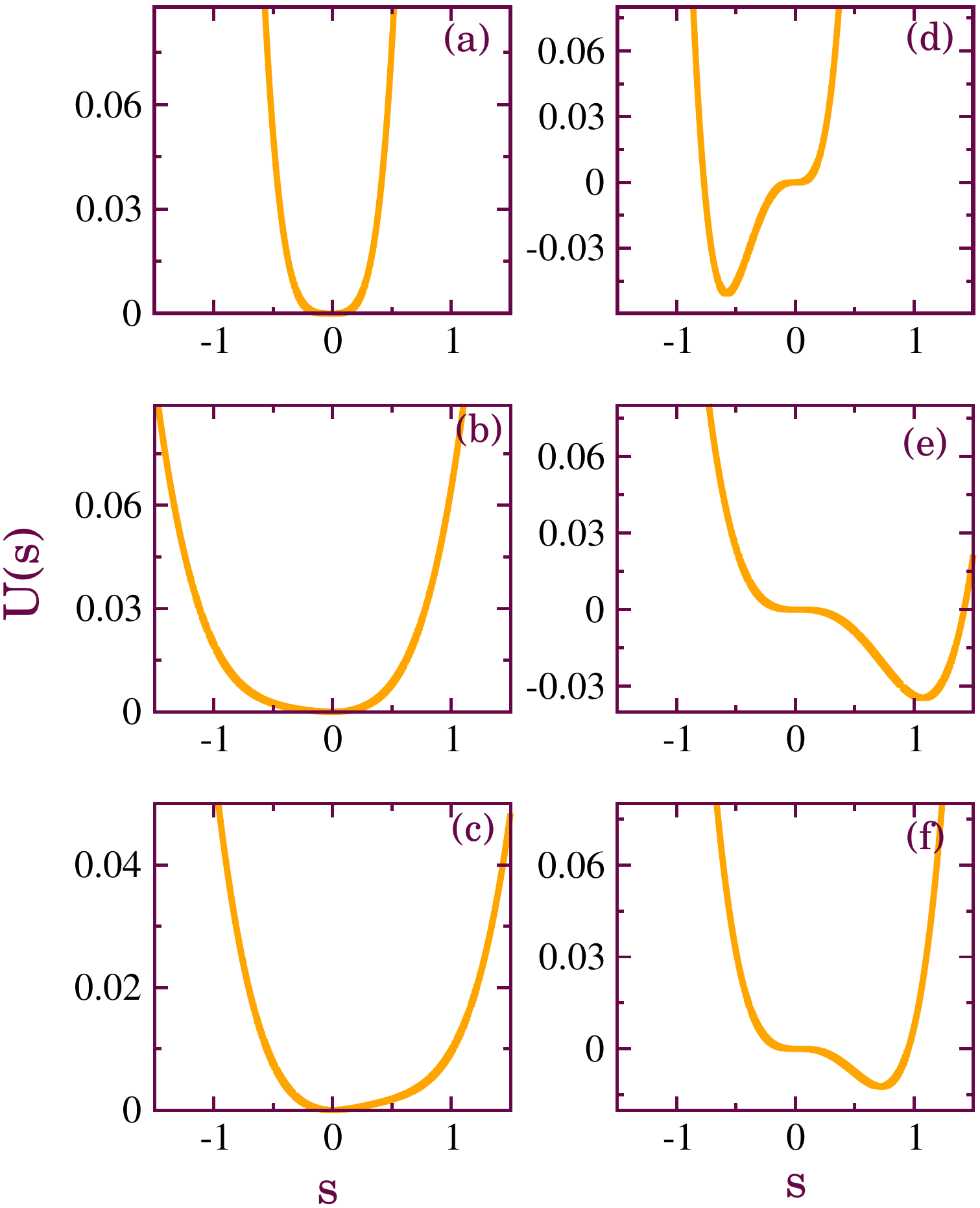}
	\caption{Typical portraits of modes from group A (panels (a), (b) and (c)) and of group
		B (panel (d), (e) and (f).)}
	\label{modes}
\end{figure}
%%%%%%%%%%%%%%%%%%%%%%%%%%%%%%%%%%%%%%%%%%%%%%%%%%%%%

{\bf Results and analysis}: We have created $6\times 10^4$ independent realizations of this glass, and after cooling each to an inherent state at $T=0$ we computed the Hessian there and diagonalized it. Having 8000 modes (eigenfunctions) for each realization (i.e. 4.8$\times 10^8$ modes all in all), we first collected those with a small
participation ratio. Here the participation ratio of the $k$th mode  $PR^{(k)}$ is defined as
\begin{equation}
PR^{(k)} \equiv  [N(\B \Psi^{(k)}\cdot\B \Psi^{(k)})^2]^{-1} \ .	
	\end{equation}
Focusing on modes whose participation ratio is smaller than 0.27 we found
 403 QLM's. Each one of these was analyzed to provide an expansion of the type of Eq.~(\ref{ass2}). To find the expansion in the direction of the eigenfunction $\B \Psi^{(k)}$ we define
\begin{eqnarray}
&&b_2^{(k)} \equiv \B \Psi^{(k)}\cdot \B H \cdot \Psi^{(k)} \ , \nonumber \\&&b_3^{(k)} \equiv \B\Psi^{(k)}_i [\frac{\partial^3 U}{\partial\B r_i\partial \B r_j\partial \B r_\ell}\B \Psi_\ell^{(k)}]\B \Psi_j^{(k)}\nonumber \\
&&  b_4^{(k)} \equiv \B\Psi^{(k)}_i [\B \Psi^{(k)}_\ell\frac{\partial^4 U}{\partial\B r_i\partial \B r_j\partial \B r_\ell\partial \B r_m}\B \Psi_m^{(k)}]\B \Psi_j^{(k)} \ ,
\label{defbk}
\end{eqnarray}
where repeated indices are summed upon. We should stress that in computing the expansion coefficients we have selected the direction of the harmonic eigenfunction as the relevant direction. It is possible that expanding in another direction can lead to a different quartic structure, revealing other features of the energy landscape \cite{15GL,20KRZ}. We choose the direction of $\B \Psi^{(k)}$ since we want $b_2^{(k)}$ to be the frequency of our QLM, the one that contributes to the density of states leading to Eq.~(\ref{dof}). Excitations in other directions will lead to a response with a host of frequencies.

Having computed the coefficients, we separated our
403 QLM's into two group, 278 of them satisfying the inequality (\ref{ineq}), and 125 for which $|b^{(k)}_3|> \sqrt{3b^{(k)}_2b^{(k)}_4}$. It turns out that not even one of the modes that satisfy the 
constraint (\ref{ineq}) has a double-well structure. Denoting the group of modes that satisfy Eq.~(\ref{ineq}) as group A and those that do not as group B, we show in Fig.~\ref{modes} typical portraits of the quartic expansion of both groups.

\begin{figure}
	\includegraphics[width=0.40\textwidth]{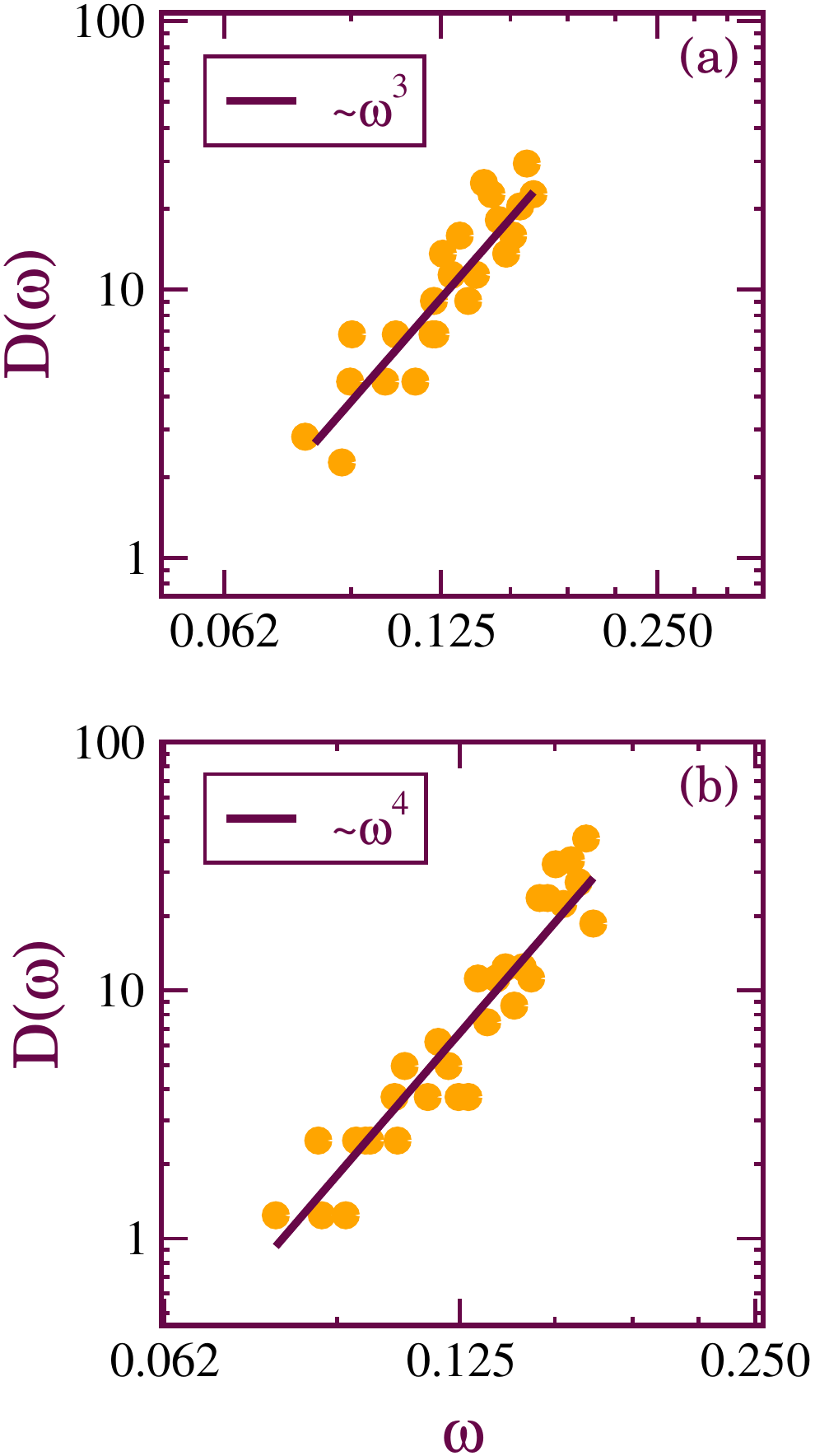}
	\caption{Density of states of modes from group B (panel (a) and of group
		A (panel (b).}
	\label{densities}
\end{figure}

According to the analytic discussion presented above we expect that the density of states of modes of group A will obey Eq.(\ref{dof}) although they have no double well structure.
On the other hand, the modes of group B that do have a double well structure are expected
to have a density of states proportional to $\omega^3$ \cite{15BMPP}. These expectations are born out by the data.  In Fig.~\ref{densities} we show the double logarithmic plots of the density of states of modes of group B in panel (a) and of group A in panel (b).
%%%%%%%%%%%%%%%%%%%%%%%%%%%%%%%%%%%%%%%%%%%%%%%%%%%%%%%%%%%
%%%%%%%%%%%%%%%%%%%%%%%%%%%%%%%%%%%%%%%%%%%%%%%%%%%%%

It is important to notice that the modes of group B are very fragile, with the minimum that is found in the numerics at $\B q_0$ being separated by a very shallow maximum from a
deeper minimum. This is the case for {\em all} the 125 modes belonging to group B. Scanning the magnitude of the barriers we find that all the 125 barriers span between 3.4$\times 10^{-9}$ and 4.0$\times 10^{-4}$. Due to these very shallow maxima we expect that the physical significance of modes of group B were quite limited. Any temperature fluctuations (or any external strain) will potentially cause a transition to the deeper minimum, adding them to modes of group A where the inequality (\ref{ineq}) is obeyed. After such a transition they would
contribute to the $\omega^4$ density of states. In contradistinction, the modes of group A are very robust, being very stable against small perturbations. The interesting point to
observe is that they have nothing to do with two-level systems or double-well potentials.
%%%%%%%%%%%%%%%%%%%%%%%%%%%%%%%%%%%%%%%%%%%%%%%%%%%%%%%%%%%
\begin{figure}
	\includegraphics[width=0.45\textwidth]{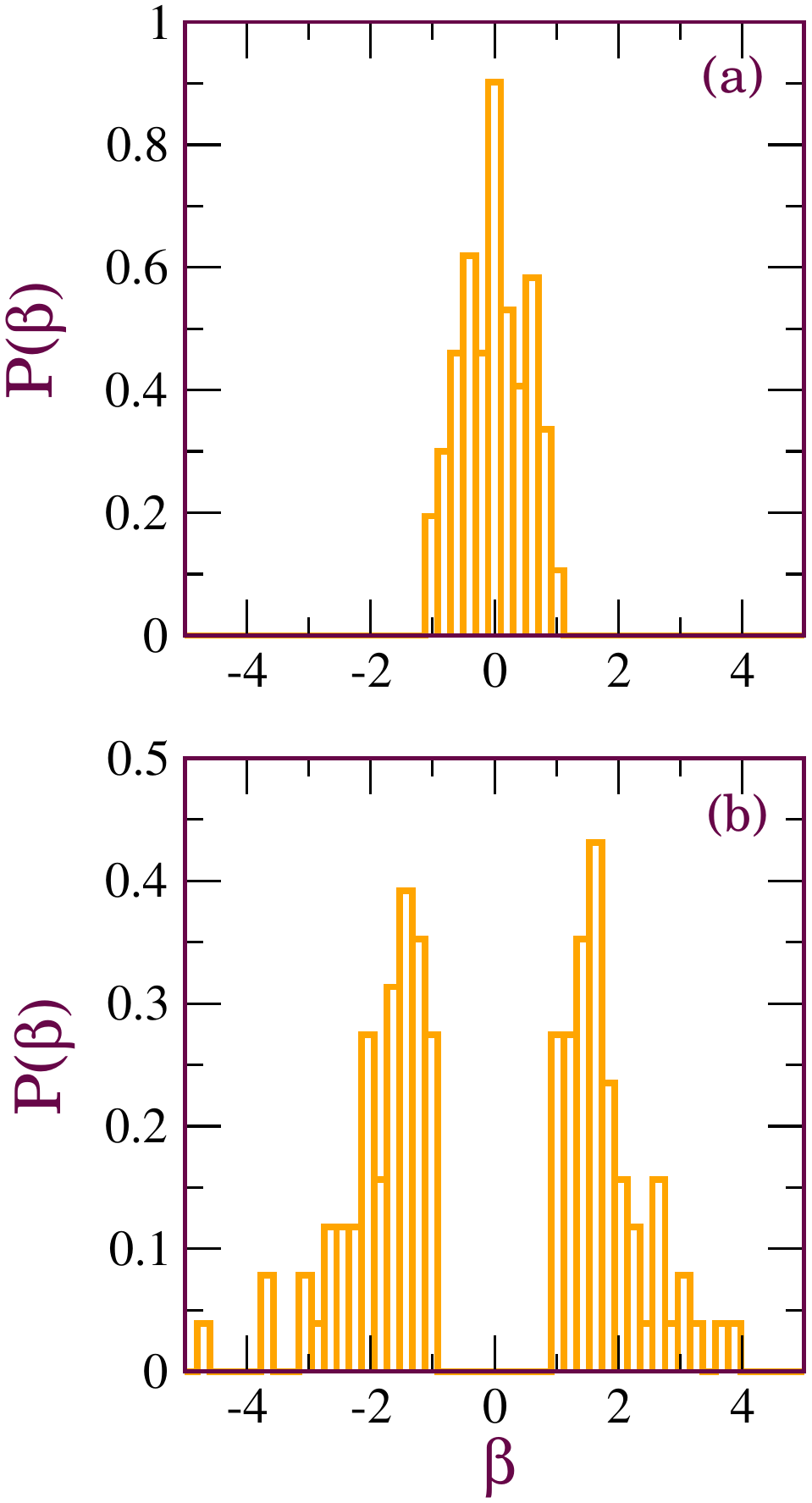}
	\caption{The PDF $P(\beta)$ for group A (panels (a)) and for group
		B (panel (b).}
	\label{pdfbk}
\end{figure}
%%%%%%%%%%%%%%%%%%%%%%%%%%%%%%%%%%%%%%%%%%%%%%%%%%%%%

\begin{figure}
	\includegraphics[width=0.45\textwidth]{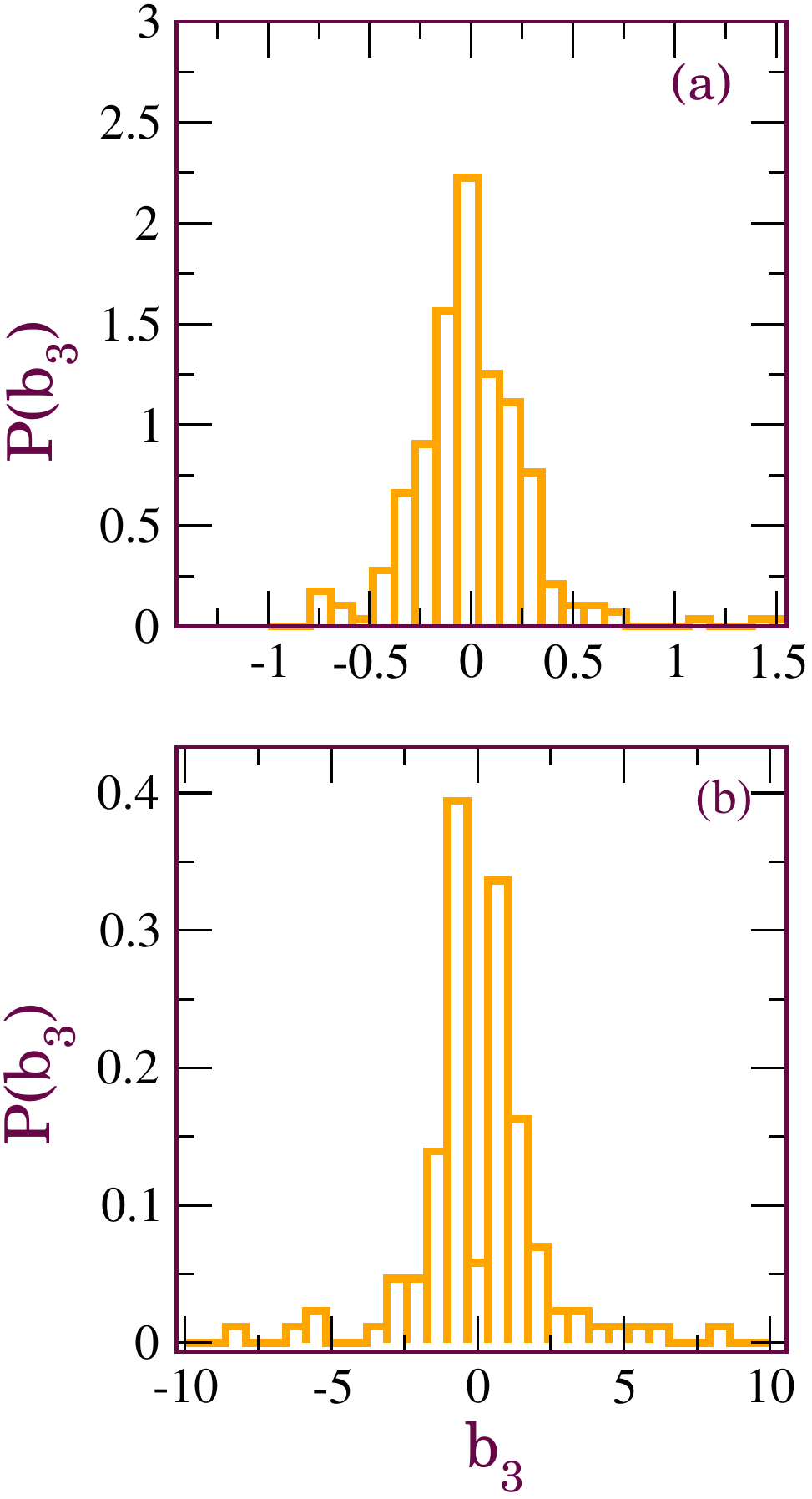}
	\caption{The PDF $P(b^{(k)}_3)$ for group A (panels (a)) and for group
		B (panel (b).}
	\label{bk3}
\end{figure}

We should examine which of the coefficients in the expansion (\ref{defbk}) is mostly responsible for the split between the two groups A and B of QLM's. To this aim we define the ratio $\beta^{(k)}\equiv b^{(k)}_3/\sqrt{3b^{(k)}_2b^{(k)}_4}$ and present in
Fig.~\ref{pdfbk} the PDF's $P(\beta^{(k)})$, where $k$ is taken from all the modes of both groups. Clearly, for group B $\beta^{(k)}$ is excluded from small values, being by definition larger than
unity . The question remains however which of the three coefficients, $b^{(k)}_2$ $b^{(k)}_3$ or $b^{(k)}_4$ is mostly responsible for the separation to the two groups.
Close examination shows that $b^{(k)}_2$ and $b^{(k)}_4$ span the same range in groups A and B. It is $b^{(k)}_3$ which differs greatly. In Fig.~\ref{bk3} we show the PDF's of the cubic coefficient. The PDF of group A peaks at very small values around zero, whereas for group B it has a dip at small values, and the distribution includes much larger (an order of magnitude large) values than in group A. We conclude that it is the cubic term that determines
whether a given QLM belongs to group A or B in the present case. Of course the existence of a large value of $b^{(k)}_3$ leads to the existence of the deeper minimum away from $\B q_0$
in all the profiles of the modes belonging to group B.  
%%%%%%%%%%%%%%%%%%%%%%%%%%%%%%%%%%%%%%%%%%%%%%%%%%%%%%%%%%%
%%%%%%%%%%%%%%%%%%%%%%%%%%%%%%%%%%%%%%%%%%%%%%%%%%%%%
We should stress at this point that in the present study all the coefficients that are quoted are not taken from a presumed distribution or a random matrix theory. They are all computed explicitly for the chosen model of glass former, using its actual Hamiltonian and the modes of the Hessian. They represent the actual microscopic environment of the glass, being a true realization of the physical nature of QLM's computed from scratch. Thus the lack of two-level systems should be taken as an indication of the generic physics of ultra-low temperature glasses. We also mention in passing that the coefficients
of the expansion are not randomly distributed. What is the precise source of this lack of randomness is an interesting question for future research, as it appears to be at the basis of glassy randomness. Although amorphous, glassy solids do not provide license for any random PDF. Correlations exist and need to be taken into account. 

As said in the introduction, in spite of the time-honored belief in two-level systems for the study of low-temperature physics of glassy solids, not everybody agreed \cite{13LV,19BZ}. Yu and Legget \cite{88YL} for example wrote quite explicitly: ``what we are disputing is the claim that the TLS model...is the unique and universal explanation of the behavior, in particular...of amorphous solids below 1K." As a conclusion of the present study we concur. At least as far as the universal density of states of QLM's Eq.~(\ref{dof}) is concerned, we have demonstrated by explicit enumeration of the modes, that those giving rise to this density of states are robust, single-minimum states,
that have nothing to do with two-level systems. Moreover, those modes that do show two minima
are all very fragile, and if they would remain relevant in any physical context, their density of states will go like $\omega^3$ rather than $\omega^4$.

{\bf Acknowledgments}:  We thank Edan Lerner for useful discussions and comments. This work has been supported in part by the Minerva Foundation, Munich, Germany, and by the US-Israel Binational Science Foundation.

\bibliography{biblio}
\end{document}